# Ultrafast Optical Switching Using Photonic Molecules in Photonic Crystal Waveguides

Yanhui Zhao, Chenjiang Qian, Kangsheng Qiu, Yunan Gao and Xiulai Xu[*]

*Beijing National Laboratory for Condensed Matter Physics, Institute of Physics,*
*Chinese Academy of Sciences, Beijing 100190, People's Republic of China*

[*]xlxu@iphy.ac.cn

**Abstract:** We study the coupling between photonic molecules and waveguides in photonic crystal slab structures using finite-difference time-domain method and coupled mode theory. In a photonic molecule with two cavities, the coupling of cavity modes results in two super-modes with symmetric and anti-symmetric field distributions. When two super-modes are excited simultaneously, the energy of electric field oscillates between the two cavities. To excite and probe the energy oscillation, we integrate photonic molecule with two photonic crystal waveguides. In coupled structure, we find that the quality factors of two super-modes might be different because of different field distributions of super-modes. After optimizing the radii of air holes between two cavities of photonic molecule, nearly equal quality factors of two super-modes are achieved, and coupling strengths between the waveguide modes and two super-modes are almost the same. In this case, complete energy oscillations between two cavities can be obtained with a pumping source in one waveguide, which can be read out by another waveguide. Finally, we demonstrate that the designed structure can be used for ultrafast optical switching with a time scale of a few picoseconds.

## 1. Introduction

To implement quantum information processing, devices such as single qubits [1, 2] single-photon sources [3-6] and single-photon detectors [7-9] have been fabricated in recent 20 years. However, to achieve quantum photonic network using these devices [10, 11] is still a major challenge. Coupling quantum nodes with quantum channels is a key step to realize photonic network. The quantum nodes and quantum channels can be implemented with optical microcavities [12] and waveguides. Photonic crystal

(PhC) cavities have been widely used as quantum nodes because of their high quality factors and small mode volumes [13-15], and PhC waveguides can guide light with very low losses in plane [16, 17]. Up to now, integration of PhC cavities and waveguides has been widely investigated to achieve photonic circuits [18-22]. Studies have been carried out on the coupling between waveguides and a single cavity [23-25] or two coupled cavities [26,27]. For example, PhC channel drop filters have been designed with coupled structures consisting of two waveguides and two coupled single-mode microcavities [26]. All-optical switching based on coupled PhC nanocavities has been demonstrated using Fano resonance [27]. In this work, we report on the energy oscillation between two strongly coupled PhC cavities, i.e. photonic molecules, and their integration with two waveguides for ultra-fast optical switching.

Photonic molecules (PM) are formed by two or more coupled optical cavities. The coupling of optical modes of individual cavities in a PM generates a mode splitting [28]. In a PM with two cavities, linear superposition of the same mode of two isolated cavities results in a symmetric (S) and an anti-symmetric (AS) super-modes [29]. The interference between S and AS super-modes induces an energy oscillation between two cavities [30]. To excite and read out the energy oscillation, in-plane PhC waveguides have natural advantages as these can be easily coupled with PMs. Therefore, waveguides can be used as signal input and output in PM-based photonic network. In this paper, we study the coupling between a PM and two waveguides in PhC with finite-difference time-domain (FDTD) method [31] and coupled mode theory [32]. By optimizing the quality factors of PM super-modes, a complete energy oscillation between two cavities is obtained with a pumping source in one waveguide, and the oscillating signal can be read out by another waveguide.

## 2. Energy oscillations in a PM

*2.1 A PM with two PhC cavities*

Firstly, we consider the coupling between two L3 cavities in a PhC slab with a triangular lattice of air holes. The slab thickness is *a*, where *a* is the lattice constant, and the dielectric constant of the slab is 12.96. The air hole radius is 0.3*a*. A L3 cavity is formed by removing three in-line neighboring air holes, which has been demonstrated with high quality factors and small modal volumes [13]. Placing two L3 cavities in line with two air hole barriers [as shown in Fig. 1(a)], a PM is formed due to the coupling between evanescent optical cavity modes of two individual L3 cavities.

To suppress the leakage out of plane, two air holes at the edges of PM in the x direction are shifted away from the center by 0.15*a*. There are 35 lattice units in both x and y directions. In the FDTD simulation, a spatial resolution of 16 pixels per lattice constant has been used. A subpixel smoothing feature [31] has been employed, so that the resolution is enough to resolve the structure modification in this work. This PhC structure has photonic band gaps in the TE-like modes with electric field in plane and magnetic field out of plane. The first photonic band gap is between the normalized frequencies of 0.22328 and 0.29659 (*a/λ*), where *λ* is the wavelength of light in vacuum.

To analyze the eigen frequencies of the coupled system, we define $a_1$ and $a_2$ as the field amplitudes of two individual cavities. With coupled mode theory we obtain:

$$\frac{da_1}{dt} = -i\omega_1 a_1 - \frac{\gamma_1}{2} a_1 - i\kappa a_2 \qquad (1)$$
$$\frac{da_2}{dt} = -i\omega_2 a_2 - \frac{\gamma_2}{2} a_2 - i\kappa a_1$$

where $\omega_1$ and $\omega_2$ are the resonance frequencies of the two cavities. $\gamma_1$ and $\gamma_2$ denote the losse rates of the cavities. $\kappa$ represents the coupling between two cavities. The complex eigen angular frequencies are

$$\omega_\pm = \frac{1}{2}\left[\omega_1 + \omega_2 - i\frac{\gamma_1 + \gamma_2}{2}\right] \pm \frac{1}{2}\sqrt{\left[(\omega_1 - \omega_2) - \frac{i}{2}(\gamma_1 - \gamma_2)\right]^2 + 4\kappa^2} \qquad (2)$$

This result indicates that the coupling between two cavities generates a splitting in real frequencies. And if the square root part in Eq. (2) gives a complex number, the imaginary parts of eigen frequencies are also split.

The spectrum of the coupled system in Fig. 1(a) has been calculated using FDTD method. When two identical L3 cavities are brought together with a two-air-hole barrier, the coupling of fundamental resonant modes of individual cavities generates two super-modes, which can be verified by the split peaks in Fig. 1(b). The eigen frequencies of two super-modes are 0.23248 and 0.23281 (*a/λ*). The different linewidths of two peaks indicate the splitting of imaginary parts of eigen frequencies. We then calculate the field distributions and quality factors corresponding to the split eigen frequencies. In TE-like modes of cavities, electric field is the superposition of $E_x$ and $E_y$ components, while the magnetic field has only z component ($H_z$). To indicate super-modes clearly, we calculate $H_z$ field distribution. Figure 1(c) and 1(d) show the spatial distributions of magnetic field component ($H_z$) of two super-modes. The field distributions with eigen frequency 0.23248 (*a/λ*) [Fig. 1(c)] show an odd parity along the x direction, whereas eigen frequency 0.23281 (*a/λ*) [Fig. 1(d)] has an even parity. The two super-modes are denoted by anti-symmetric (AS) and symmetric (S) modes respectively, corresponding to their odd or even field parities. The quality factors of AS and S modes are 7880 and 33530, respectively. To explain the difference of the quality factors of AS and S modes, we employ the vertical total internal reflection mechanism [13]. The wave vector distribution of cavity mode can be obtained from the Fourier transformation of the field spatial distribution of that mode. Light with wave vector inside light cone is not confined by total internal reflection, and will leak out of the cavity. That outside of light cone can be localized in cavity. Due to different field distributions of AS and S modes, the wave vector distributions are also different, resulting in a different integral intensity in the light cone. Larger the integral intensity in the light cone, higher the leakage out of cavity and lower the quality factor, vice versa.

We then analyzed energy time-evolution in two coupled cavities. In Fig. 1(a), the two coupled L3 cavities are identical. So both the resonant frequencies and cavity loss rates are same, i.e. $\omega_1 = \omega_2 = \omega_0$, $\gamma_1 = \gamma_2 = \gamma_0$. We consider the field initially localized at cavity 1 ($a_1(0)=1$, $a_2(0)=0$). With Eq. (1) and (2), the time evolution of the fields at two cavities can be calculated as

$$a_1(t) = \frac{1}{2i} e^{-i\omega_+ t} \left[ 1 + e^{i(\omega_+ - \omega_-)t} \right]$$
$$a_2(t) = \frac{1}{2} e^{-i\omega_+ t} \left[ 1 - e^{i(\omega_+ - \omega_-)t} \right]$$
(3)

where $\omega_+$ and $\omega_-$ are defined as $\omega_+ = \omega_S - i\gamma_S$ and $\omega_- = \omega_{AS} - i\gamma_{AS}$. $\omega_S$ and $\omega_{AS}$ are the eigen angular frequencies of the two super-modes. $\gamma_S$ and $\gamma_{AS}$ are the loss rates of S and AS modes respectively. Then the field intensity in each cavity can be written as

$$|a_1(t)|^2 = \frac{1}{4} e^{-2\gamma_S t} \left| 1 + e^{i(\omega_S - \omega_{AS})t} e^{-(\gamma_{AS} - \gamma_S)t} \right|^2$$
$$|a_2(t)|^2 = \frac{1}{4} e^{-2\gamma_S t} \left| 1 - e^{i(\omega_S - \omega_{AS})t} e^{-(\gamma_{AS} - \gamma_S)t} \right|^2$$
(4)

With the parameters extracted by FDTD simulation in the case of Fig. 1, the loss rates $\gamma_S$ and $\gamma_{AS}$ can be calculated using $Q_S = \omega_S/(2\gamma_S)$ and $Q_{AS} = \omega_{AS}/(2\gamma_{AS})$. The time evolution of field intensity in two individual cavities can be calculated using Eq. (4). Figure 2(a) demonstrates that energy oscillates between the two cavities. However, the energy oscillations are not complete, which means the valleys of oscillations do not approach zero intensity. The incomplete oscillation indicates that light does not transfer fully from one cavity to another. From Eq. (4), it can be seen that the difference between $\gamma_S$ and $\gamma_{AS}$ determines whether the energy oscillations are complete or not.

Figure 2(b) shows FDTD results of energy oscillations in the structure of Fig 1(a) with a TE mode Gaussian source located at cavity 1. The center frequency of Gaussian source has been set at the center of eigen frequencies of AS and S modes, i.e. 0.23265 (*a/λ*). The frequency range of Gaussian source covers both AS and S modes with a frequency width of 0.01 (*a/λ*). The time evolution of electric field energy has been monitored at the centers of each cavity. It can be seen that the energies oscillate periodically between two cavities and the oscillations are not complete, which is in good agreement with coupled mode theory. It should be noted that there is time delay in oscillations in the case of the FDTD simulation, comparing with coupled mode theory. This is because it needs some time to excite the cavity modes with a Gaussian source in simulation.

From field distributions in Fig. 1(c) and 1(d), we can infer that the energy oscillations are due to cancellation and enhancement of field amplitudes of AS and S modes [30]. A complete energy oscillation requires a perfect destructive interference between AS and S modes, which results from fine phase matching and comparable field amplitudes of two super-modes. From Eq. (4), complete energy oscillations can be achieved by equalizing the loss rates of AS and S modes, i.e. $\gamma_{AS} = \gamma_S$.

*2.2. Structure optimization for complete energy oscillations in a PM*

From the above discussions, complete energy oscillations between two cavities in a PM require equalized loss rates of two super-modes. We need to adjust the coupled structure in Fig. 1(a) to get same loss rates. From expressions, $Q_{AS} = \omega_{AS}/(2\gamma_{AS})$ and $Q_S = \omega_S/(2\gamma_S)$, the same loss rates of two super-modes do not have same quality factors because of the difference between $\omega_{AS}$ and $\omega_S$. However, if the eigen frequency splitting is small and the quality factors are relatively high, the loss rates for both super-modes are proximately same by equalizing quality factors. With this approximation, we optimize the quality factors to get the same loss rates for both super-modes.

From the field distributions in Fig. 1(c) and 1(d), we can see that the fields located at air holes ($r_m$) which are surrounded by green dashed circles are different. There are fields distributed in these air holes for S mode, whereas no field observed in the case of AS mode. It can be inferred that tuning the radius $r_m$ can modify the quality factors of AS and S modes differently and effectively, which can be used to achieve equalized quality factors.

Figure 3(a) shows the eigen frequencies of AS and S modes by varying the radius $r_m$ from 0.3*a* to 0.44*a*. Eigen frequencies for both AS and S modes become higher with increasing radius $r_m$, which is due to the fact that the average refractive index of structure decreases by increasing the radius. The air hole radius has more influence on S mode eigen frequency than AS mode, because there are more field distributed in air holes of $r_m$ in S mode. So the eigen frequency splitting becomes larger as $r_m$ increases. Figure 3(b) shows the quality factors of AS and S modes as a function of $r_m$. Quality factors for two modes first increase, reach a maximum value and then decrease. Varying the air hole radius $r_m$ modifies the field spatial distributions of super-modes, thus the amount of light with wave vector inside the light cone changes accordingly. For both AS and S modes, there is an optimized air hole radius $r_m$ which can achieve least amount of light with wave vector inside the light cone, resulting in a maximum quality factor.

At $r_m$ = 0.382*a*, the quality factors of AS and S modes are nearly equal, $Q_{AS} = 33020$ and $Q_S = 32650$. The eigen angular frequencies of AS and S modes are $\omega_{AS} = 0.23286 \times 2\pi(a/\lambda)$ and $\omega_S = 0.23343 \times 2\pi(a/\lambda)$, respectively. In this case, the loss rates of two super-modes are nearly same. With these parameters given by FDTD simulations, the time evolution of field intensities can be calculated with Eq. (4). Figure 3(c) shows the theoretical results with a complete energy oscillation between two cavities. In addition, the energy evolution is also calculated by FDTD method. A Gaussian source has been set at cavity 1, with a center frequency of 0.23315 (*a/λ*) (center of the eigen frequencies of two super-modes). A frequency width of 0.01 (*a/λ*) has been used, which covers both AS and S modes. Then electric field energy has been monitored at the centers of two individual cavities. The simulated results in Fig. 3(d) show complete energy oscillations, comparing with those in Fig. 2(b). Comparing the energy evolutions in the cases of $r_m$ = 0.3*a* [Fig.

2(b)] and $r_m$ = 0.382$a$ [Fig. 3(d)], we can infer that the periods of the energy oscillations are different, which indicates the oscillation periods can be tuned. This result can also be inferred theoretically from Eq. (4).

## 3. Coupled PM-waveguide structure for ultrafast optical switching

### 3.1. Coupling between a PM and two waveguides

PhC waveguide is formed by removing one row of air holes in photonic lattice. This type of defect induces guided modes with frequencies in photonic band gap range [17, 33], which can guide light with very low losses for a wide range of frequencies [16]. The waveguide in this work is formed by removing one row of air holes along the Γ-K direction (W1 waveguide) [Fig. 4(a)]. The waveguide band diagram was calculated with a free source software MPB [34]. Figure 4(b) shows the calculated band diagram of the TE modes of W1 waveguide with same structure parameters in Fig. 1(a). The red and blue solid lines show the zeroth-order and the first-order waveguide modes, respectively. We note that only the zeroth-order mode in waveguide can spectrally match AS and S modes in PM discussed before. This spectral matching between waveguide modes and PM super-modes might induce an efficient coupling between them.

Theoretical and experimental studies have shown that an efficient coupling between waveguide and L3 cavity modes can be achieved with L3 cavity tilted to waveguide axis by an angle of 60°, in which the field overlap between evanescent cavity mode and waveguide mode is maximized [23]. We adopt this configuration for coupled PM and waveguide structure as shown in Fig. 5(a). Two L3 cavities are aligned in line along the x-axis with two-air-hole barrier. The two air holes at the edges of PM along the x-axis are shifted by 0.15$a$ away from center to suppress out-of-plane losses. Two W1 waveguides, waveguide In and waveguide Out, are tilted with respect to the x-axis by an angle of 60°. The waveguides are separated from PM by three air holes.

With waveguides coupled to a PM, the system can be described as

$$\begin{aligned}
\frac{da_1}{dt} &= -i\omega_0 a_1 - \frac{\gamma_c + \gamma_{wg}}{2} a_1 - i\kappa a_2 - \sqrt{\gamma_{wg}} a_{In} \\
\frac{da_2}{dt} &= -i\omega_0 a_2 - \frac{\gamma_c + \gamma_{wg}}{2} a_2 - i\kappa a_1 \\
a_{Out} &= \sqrt{\gamma_{wg}} a_2
\end{aligned} \qquad (5)$$

where $\omega_0$ denotes the resonant frequency of two individual cavities, $\gamma_c$ is the loss rate of individual cavity without waveguide, and $\gamma_{wg}$ corresponds to the loss rate into waveguide. The total loss rate of each cavity in coupled system is given by $\gamma_{total} = \gamma_c + \gamma_{wg}$. In this coupled system, a complete energy oscillation requires the equalization of the total loss rates of AS and S modes in the PM with considering the waveguides.

In structure of Fig. 5(a), the resonant frequencies of AS and S modes are 0.23242 and 0.23276 (*a/λ*) respectively. The quality factor of AS mode is 3800, and that of S mode is 6760. We then calculated the transmission spectrum with FDTD. A Gaussian source with frequency center 0.23259 (*a/λ*) and frequency width 0.01 (*a/λ*) has been set in waveguide In. The energy flux has been monitored at waveguide Out. The transmission spectrum shows two peaks, corresponding to AS and S modes. Figure 5(c) and 5(d) show the $H_z$ field distributions of the two super-modes.

*3.2. Structure optimization for reading out energy oscillations in a PM by a waveguide*

In order to realize complete energy oscillations, we tune $r_m$ of the air holes marked by the green dashed circles in Fig. 5(c) and 5(d) to achieve equalized quality factors using the same method in section (2.2). Figure 6(a) and 6(b) show the resonant frequencies and quality factors of AS and S modes as a function of $r_m$. Similar features of resonant frequencies and quality factors are observed as in section (2.2). At $r_m$ = 0.388*a*, the quality factors of AS and S modes are equalized ($Q_{AS} = Q_S = 6180$) with resonant frequencies at 0.23284 and 0.23344 (*a/λ*), respectively.

We have shown that complete energy oscillations require equalized total loss rates of AS and S modes when excitation source is located in one cavity. In the case of exciting PM by waveguide modes, the coupling strengths between waveguide modes and two PM super-modes are also required to be same, so that AS and S modes can be excited equally by waveguide modes. In our structure, the total quality factors of AS and S modes are equalized ($Q_{AS} = Q_S = 6180$) at $r_m$ = 0.388*a*. With $r_m$ = 0.388*a*, the quality factors of super-modes without waveguide are $Q_{c-AS} = 36700$ and $Q_{c-S} = 28450$. The quality factors due to the losses into waveguides can be calculated as $Q_{wg-AS} = 7431$ and $Q_{wg-S} = 7895$. The close quality factors indicate nearly equal coupling strengths between waveguide modes and two super-modes.

Finally, we excite the PM with waveguide modes. With optimized parameter $r_m$ = 0.388*a*, a Gaussian source with center frequency 0.23314 (*a/λ*) and frequency width 0.01 (*a/λ*) has been set in waveguide In. Figure 6(c) and 6(d) show the energy oscillations in two cavities and waveguide Out. It can be seen that complete energy oscillations have been achieved in PM and the oscillations can be read out by waveguide Out. To excite a PM with waveguide and realize complete energy oscillations in PM, two conditions are required: first, the total loss rates of super-modes of a PM should be adjusted to be equal; second, the coupling losses between waveguide and two PM super-modes need to be close. The period of the energy oscillations in Fig. 6(c) and 6(d) is about 1700 (*a/c*). If the lattice constant *a* is set as 300 nm, the period is about 1.7 ps. In the case of complete energy oscillations, the contrast is high which makes the coupled system useful for ultrafast optical switching. It should be noted that the resonant frequency splitting can be used to adjust the period on demand, larger the resonant frequency splitting faster the energy oscillations.

## 4. Conclusion

In summary, we studied the coupling between PM and waveguides in PhC slab using FDTD calculation method and coupled mode theory. We designed a coupled PM-waveguide structure in which complete energy oscillations are achieved and the oscillating information can be transmitted to waveguides. The coupling strengths of waveguide modes with two super-modes of a PM and total quality factors of two super-modes can be adjusted to nearly equal by optimizing the radius of the air holes between the two cavities. With the optimized structure, an ultrafast energy oscillation with a period of a few picoseconds is obtained, which can be used as an ultrafast optical switch.

**Acknowledgements**

This work was supported by the National Basic Research Program of China under Grant No. 2013CB328706 and 2014CB921003; the National Natural Science Foundation of China under Grant No. 91436101,11174356 and 61275060; the Strategic Priority Research Program of the Chinese Academy of Sciences under

Grant No. XDB07030200 and the Hundred Talents Program of the Chinese Academy of Sciences.

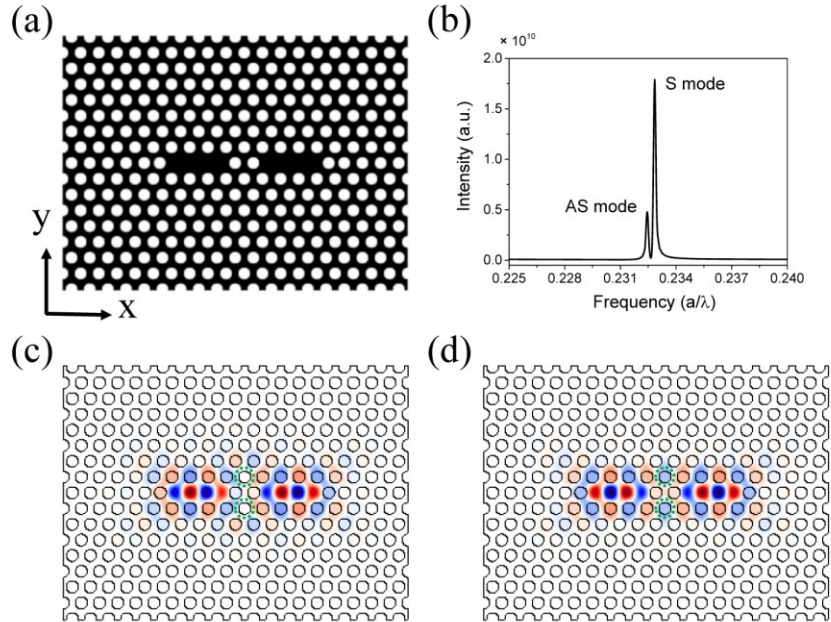

Fig. 1. A PM formed of two coupled PhC L3 cavities. (a) Two L3 cavities are aligned in line along the x-axis and separated by two air holes. The lattice constant is *a*, air hole radius is 0.3*a* and slab thickness is *a*. The dielectric constant of slab is 12.96. (b) Resonant spectrum of coupled system. AS and S mode correspond to odd and even parities of the field distributions, respectively. (c) $H_z$ field distribution of AS mode of PM. (d) $H_z$ field distribution of S mode of PM. The air holes enclosed by green dashed circles with radius $r_m$ are used to optimize the coupled structure.

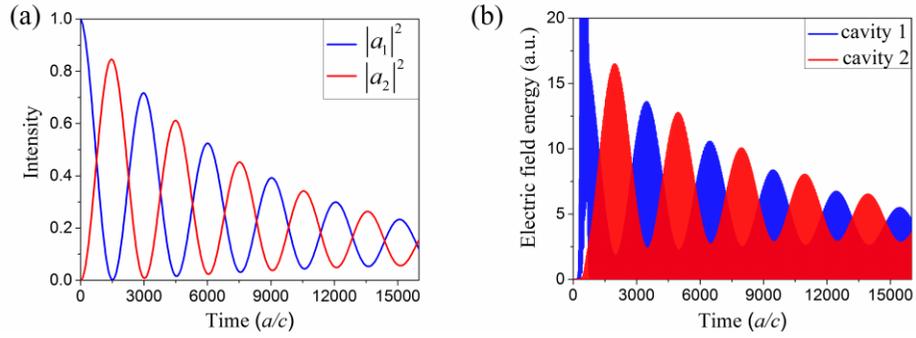

Fig. 2. Energy oscillations between two cavities of a PM. (a) Time evolution of field intensities in a coupled system within cavity 1 and cavity 2 in the case of coupled mode theory, indicating an incomplete energy oscillation between two coupled cavities. (b) FDTD simulation of time evolution of electric field energy in a PM having structure parameters in Fig. 1(a). The energy oscillations between two cavities are not complete as well, which is in a good agreement with the results in (a).

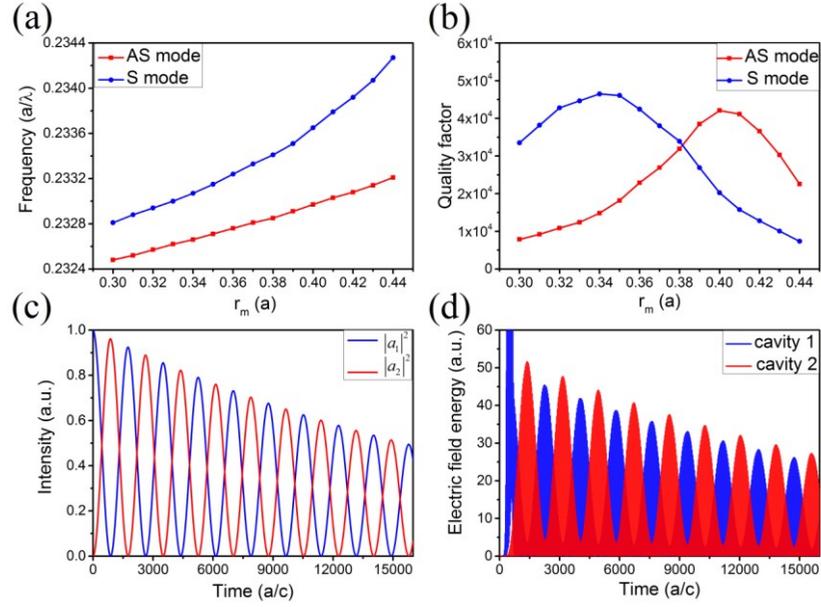

Fig. 3. Structure optimization for complete energy oscillations in a PM. (a) Eigen frequencies and (b) quality factors of AS and S modes as a function of $r_m$. (c) Energy oscillations between two cavities by coupled mode theory when the quality factors of AS and S modes are equalized at $r_m=0.382a$. (d) FDTD simulation results with same parameters as used in the coupled mode theory.

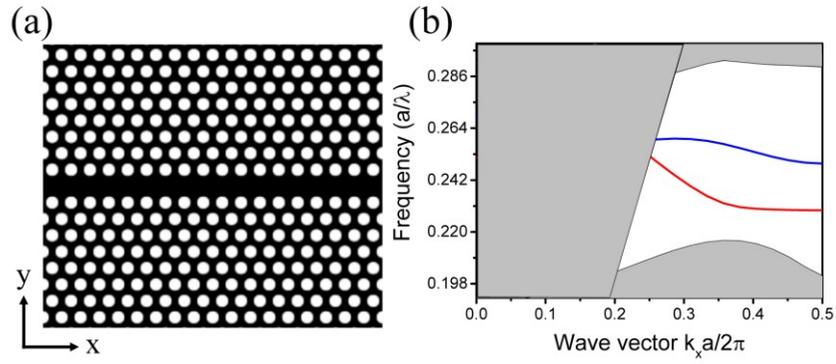

Fig. 4. Photonic crystal waveguide and the photonic band diagram. (a) A W1 waveguide structure formed by removing one array of air holes in PhC structure. (b) Band diagram of W1 waveguide. The red and blue solid lines show the zeroth-order and the first-order waveguide modes.

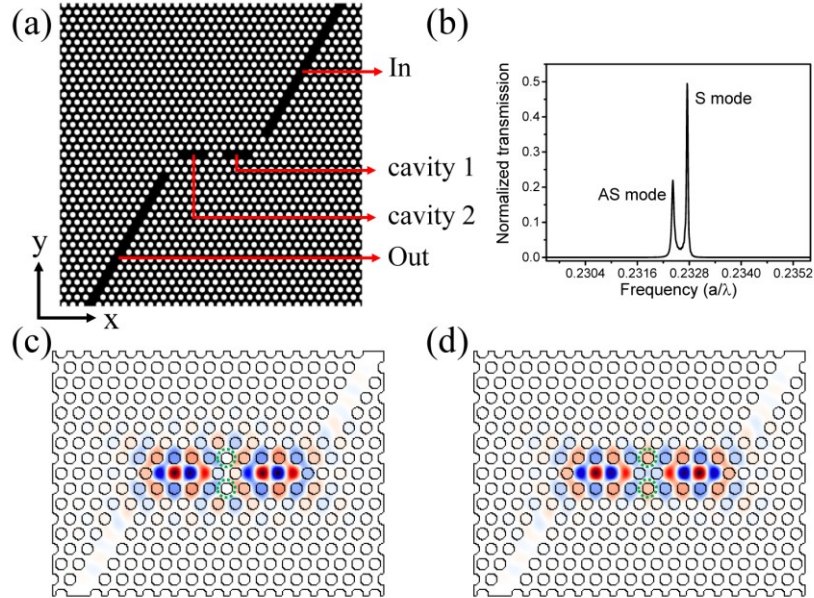

Fig. 5. Coupling between a PM and waveguides. (a) Coupled structure between a PM and two waveguides. PM is aligned along the x-axis and two waveguides are tilted with respect to the x-axis by 60° angle. The PM and the waveguides are separated by three air holes. Waveguide In and waveguide Out are used as signal input and output respectively. (b) Transmission spectrum of the coupled structure. A Gaussian source is set in waveguide In and energy flux is detected at waveguide Out. Two peaks correspond to the split AS and S modes. (c) and (d) show $H_z$ field distributions of AS and S modes, respectively.

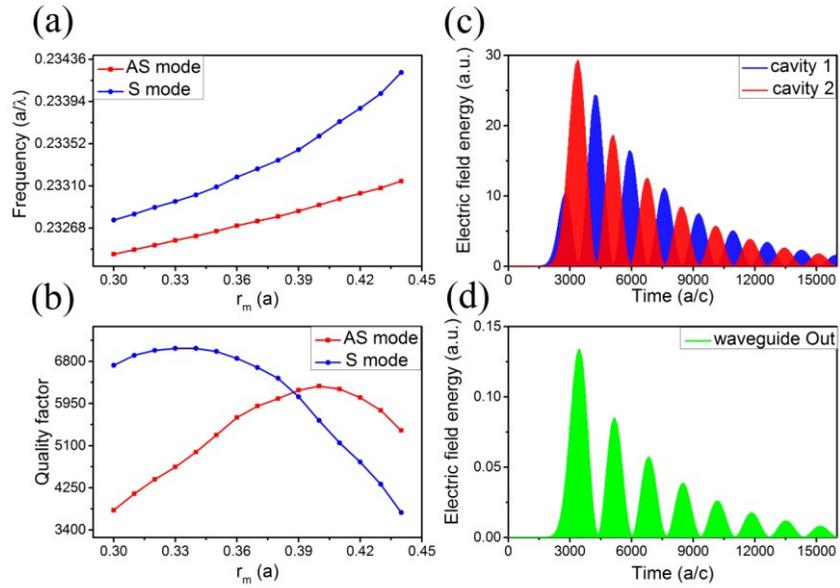

Fig. 6. Structure optimization for complete energy oscillations in a coupled PM-waveguide structure. Eigen frequencies (a) and quality factors (b) of AS and S modes as a function of $r_m$. (c) The energy oscillations between two cavities when the PM is excited by waveguide modes. (d) The energy oscillations can be read out by another waveguide.